# A novel variable-distance antenna test range and high spatial resolution corroboration of the inverse square law for 433.5 MHz radiation

Christoph de Haën[(1)°], Giancarlo Baldini[(2)] and Matthias Erhardt[(3)]

[(1)] *Thalwil, Switzerland. E-mail: cdehaen@bluewin.ch*

[(2)] *Zurich, Switzerland. E-mail: gbaldini@sunrise.ch*

[(3)] *Lyss, Switzerland. E-mail: erhardt.lyss@besonet.ch*

A novel, low-budget, open-air, slant-geometry antenna test range for UHF radiation is presented. It was designed primarily to facilitate variation of the distance between emitter and receiver antennas, but has also the potential for adaptation to simultaneous variation of distance and receiver antenna orientation. In support of the validity of the range the inverse square law for 433.5 MHz radiation between two naked half-wave dipole antennas was tested with high spatial resolution from close to the far field limit outward to 46 wavelengths. The ratio of sine-amplitude input voltages to the receiver antenna at two distances between the antennas diminished in proportion to the corresponding inverse distance ratio to the power $0.9970 \pm 0.0051$ ($R^2 = 0.992$). This value is indistinguishable from the theoretical value of 1 and confirms the proportionality of the electric field strength to the inverse distance from the radiation source. Given the known proportionality of irradiance to the square of the electric field strength, the result corroborates the inverse square law for irradiance at the lowest frequency for which thus far data have been published.

**I. INTRODUCTION**

The irradiance ($Wm^{-2}$) from a point source of isotropic radiation diminishes for purely geometric reasons in proportion to the inverse square of the distance — the inverse square law. Extended emitters and those producing polarized radiation can at best approximate the isotropic emitter. Electromagnetic theory predicts that under free-space conditions the inverse square law describes well the case of an individual half-wave dipole antenna, an extended emitter whose radiation is linearly polarized. This holds for any frequency and direction, provided the measurements are made at sufficiently large distances from the dipole

---

° Corresponding author.

center. The electric field strength of a half-wave dipole, being proportional to the square root of the irradiance, diminishes in proportion to the inverse distance. The two distance dependences will be subsumed here under the common term inverse square law.

Although a plethora of studies of radio-waves in natural environments with their absorptions, reflections and scatterings have been performed,[1] to the best of our knowledge no test of the inverse square law for narrow bandwidth UHF or lower frequency radiation have been published. The inverse square law in these cases rests on a plausible extrapolation from studies at higher frequencies, supported by a common theory and successful incorporation into descriptions of complex real situations.

For the characterization of antennas at frequencies between 100 and 1000 MHz far-field antenna test ranges are preferred over near-field ranges,[2] but for such frequencies free-space conditions are notoriously difficult to simulate.[3] Ground reflection ranges or aircrafts have been employed. In the former case variation of the distance between antennas is cumbersome and variable-height towers used offer only small variations. The application of aircrafts is extremely complex. The present project aimed at developing with modest means an outdoor, variable-distance, and high-spatial-resolution antenna test range for UHF radiation. The inverse square law was tested in order to validate the range, while simultaneously filling in a lacuna in published studies at such low frequencies.

## II. VARIABLE-DISTANCE ANTENNA TEST RANGE

**Principles** — For the design of the new test range the fixed distance slant-geometry range, which has been recommended for the study of antenna orientation,[4,5] served as point of departure. The half-wave dipole emitter antenna (E) and receiver antenna (R), as well as the bandpass filter - receiving amplifier combination (A), were mounted through positioners (P) on a carrying rope (C) that lead from a wooden viewing tower to the ground at an inclination of 45° (Fig. 1). Both antenna dipoles formed right angles with C in the vertical plane through C. While E remained stationary near the ground, R and associated A were displaceable along C with the help of a system of hauling ropes ($H_1$ and $H_2$), deflection pulleys ($D_{1\text{ to }5}$) and sand bottle weights ($W_1$ and $W_2$) that balanced the forces on the R-A combination and put $H_2$ under position-independent tension. Displacement of R could be measured through an indicator (I) on $H_2$ moving along a measuring bar (B).

The following ideal design principles had to be considered. The tower should be metal-free. The environment should be free of reflecting objects above ground. Environmental radiation with frequencies detectable by the R-A combination should be absent. The terrain needs to be flat, horizontal and dry. In order to assure parallelity of E and R in any position of R, a constant incline of C has to be assured. To this effect C has to be a strongly tensioned non-metallic rope. The weights of R and A need to be kept small. Accuracy of position measurements and avoidance of hysteresis by $H_2$ owing to forth and back movement of R requires the rope to possess low elasticity and to be under constant tension. Frictional resistances to the movement of R and A should be minimized. Movement of R should be possible over a large range with minimal intervention on $W_1$ and $W_2$ and on the combination of B and I. Positioners for E and R, as well as deviation pulleys, should be metal-free.

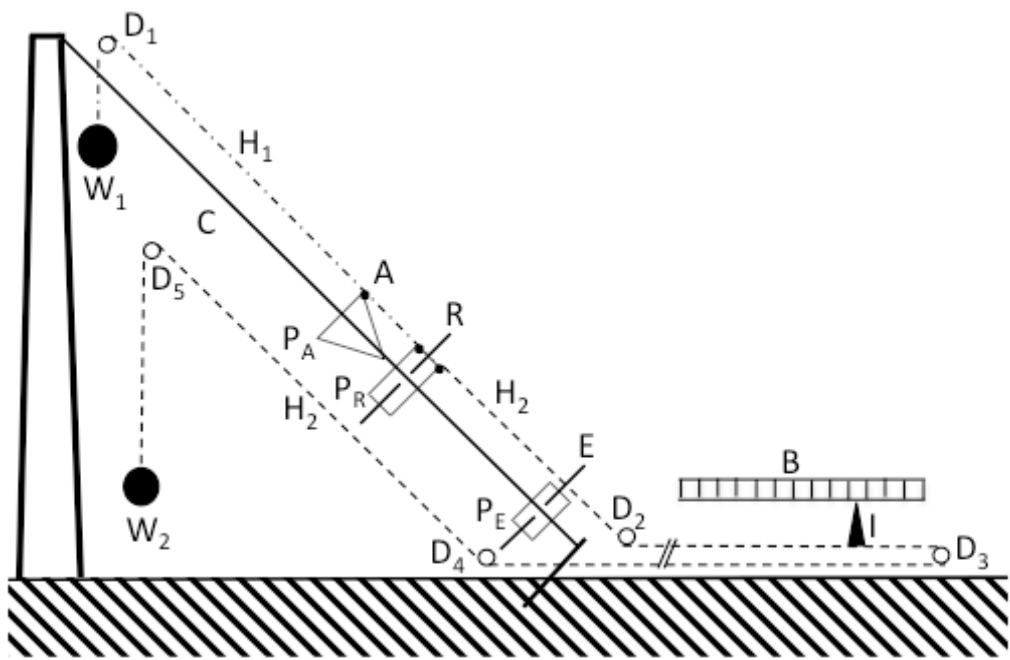

Fig. 1: Not to scale scheme of antenna test range

Reflection of radiation by A and its positioner in the direction along C should be minimized. The radiation by E should show a small bandwidth and A should possess a narrow bandpass filter. Power supplies to E and A should be stable. Experimenters, B and ancillary tools on the ground need to be at a distance from E that avoids interference with emission by E. Realization of a valid low-budget test range called for judicious concessions regarding these ideal design principles.

**Mechanics** — The test range was realized at the viewing tower, Lysserturm in Switzerland.[6] For experimentation, inclusive installation and removal of equipment one day had to suffice. Except for metallic stairs, diagonal cross bracings and connectors between logs, the tower was built in wood. A wide swath in the thin forest surrounding the tower, free of metal objects, but with some dry logs, accommodated the measuring stretch. From the bottom of the tower to that of C the flat, gravely terrain sloped upwards by 1°. The ultra-high molecular weight polyethylene rope, C, had the following properties: diameter 3 mm; breaking load 9500 N; nominal operating elongation < 1%; weight 0.045 N/m, (Liros-D-Pro R150503J, Liros GmbH, Berg, Germany). An 8 mm polyester rope and a short, mostly plastic block and tackle of mechanical advantage four, fastened C to the tower at the elevation of 34.65 m. Positioners with sliding tubes $P_A$, $P_R$ and $P_E$, threaded onto C, allowed movement of respectively A, R and E along C. At a horizontal distance of 35 m from the tower, the loaded C was tied, about 10 cm above ground, to a wooden three-membered earth anchor pile group. The tension on C was increased to 2900 N. Tape fixed $P_E$ to C, with the center of E 70 cm above ground. The idealized configuration in a slant antenna range requires the free-space radiation-pattern maximum of E to point at the center of R and its null at the specular reflection point on the ground.[4,5] Geometry allows these conditions to be met only approximately. The literature offers different instructions on where a 45° angle should be achieved.[4,5] The slope of C at the bottom was reduced by pertinent loads in various positions by no more than 1°. Based on these facts, and considering the slope of the terrain, an angle of 45 ± 1° between C and a plumb line was chosen.

Dipole antennas E and R were mounted symmetrically and parallel to an edge onto one side of their respective P, each consisting of a polymethylmethacrylate (PMMA) plate (10 cm × 10 cm × 0.5 cm). To the other side of each plate and perpendicular to the antenna, plastic sliding tubes (inner diameter 6 mm) were attached. In the case of $P_E$ the 24 cm tube was placed symmetrically and it traversed the core of a balun contiguous with the plate. The similarly affixed tube of $P_R$ measured 100 cm, 64.5 cm of which pointed upwards. In order to assure the orientation of R perpendicular to and in the vertical plane through C, $P_R$ received a verticality adjuster. It consisted in a polyvinylchloride tube (90 cm × 2.5 cm), one end of which was attached to the sliding tube side of the plate through an all-plastic twisted-plate universal joint, for a total weight of 2.3 N. The joint allowed the adjuster to swing freely in the vertical plane through C. In the vertical plane perpendicular to the former plane, the adjuster could be fixed in the orientation necessary for the correct direction of the dipole by tightening the screw of the appropriate axle. Although the $P_E$ was similarly equipped, the low elevation of E actually permitted its orientation to be fine-tuned with the adjuster sideways touching the ground. $P_A$ had to position A at a fixed distance behind R. In addition, by absorption and deviation it had to protect R and E from experimental radiation reflected by A or the ground. It consisted of three 10 cm ×10 cm × 0.5 cm brass slabs soldered to form the corner of a cube pointing along C at R (Fig. 2). In the place of the cube's space diagonal and just piercing the corner from the inside, was soldered a brass sliding tube with inner diameter 0.6 cm and length 35 cm. Along the back end of the tube was soldered an equilateral brass triangle, 7 cm on the side. It pointed vertically down when two faces of the cube corner pointed symmetrically sideward down. A cable conduit (60 cm, 0.60 N) hanging on an axle through the triangle, and leading inside a telephone cable vertically to the ground, acted as verticality adjuster for $P_A$. A 5 mm hole in the upper slab let a coaxial cable from R to A pass. Flush nickel-zinc ferrite absorber tiles (10 cm × 10 cm × 0.6 cm, Ferroxcube 4S60

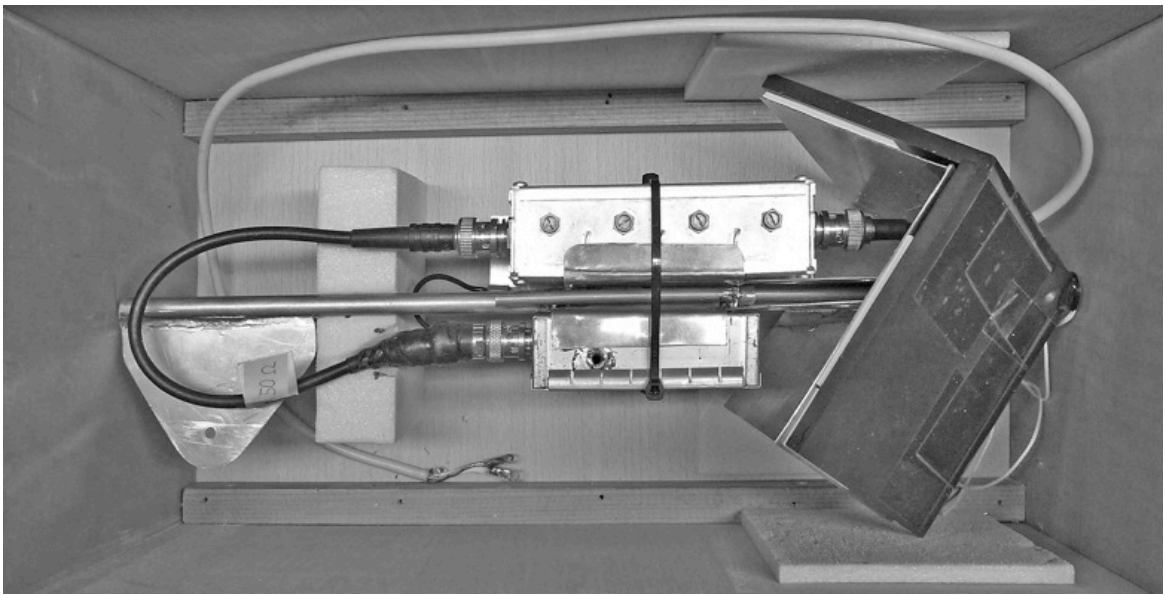

Fig. 2: Positioner ($P_A$) with filter–amplifier combination (A), packed in a box, illustrates the geometry of cube corner and triangle for cable conduit attachment. The white coaxial cable destined to connect to the receiver (R), which here is shown folded back, traverses a cube corner plate to reach the filter. The black coaxial cable connects the filter to the amplifier.

Megatron AG, Kaltbrunn, Switzerland),[7] covered the outside faces of the brass slabs. Openings that traversed the ferrite tiles corresponding to the ones in the brass slabs were drilled with diamond tools under water. Fixation of the tiles to the brass plates via two layers of double-sided polyurethane foam adhesive tape produced a dielectric layer of 3 mm. In order to avoid damage to C by sharp edges and the slightly irregular opening in the ferrite cube corner, the opening was plugged with a plastic rivet with a 6 mm axial bore. A string between the $P_A$ and $P_R$ restricted the distance from rivet to the center of R to 65 cm. Thereby the 70 cm coaxial cable between the two remained loose and did not transmit a torque. With cable ties A was attached to the brass tube in a configuration completely in the light shadow of the $P_A$. The assembly of R, $P_R$ and verticality adjuster weighed 3.8 N, while that of A, $P_A$ and stabilizer, but without telephone cable, weighed 18.3 N, for a total weight of 22.1 N.

A system of hauling ropes, $H_1$ and $H_2$, all-plastic deflection pulleys (D) and weights (W) allowed R together with A to be moved reproducibly along C while measuring its position (Fig. 1). The rope $H_1$ (1 mm) lead from $P_A$ over $D_1$, affixed to C almost at its top, to $W_1$, which weighed 30.8 N. The 1 mm measuring rope, $H_2$, was a thinner version of C and also possessed a <1% nominal operating elongation (Liros-D-Pro R150501). Without sacrificing the principles schematized in Fig. 1, the following deviations therefrom characterized the equipment actually realized. Forced by the terrain, B had to be laterally displaced from the anchoring of C by 4.6 m, but still parallel to C. This required deviation of $H_2$ by deflection pulleys affixed in torque-avoiding manner to wooden stakes. Attached to $P_R$, $H_2$ paralleled C freely passing $P_E$, followed along B and returned to deflection pulley $D_5$, after which it reached the weight $W_2$ of 15.1 N. From the location of B an operator could move R along C. When R was in its lowest position, $W_1$ reached the highest and $W_2$ the lowest elevation. This gave a measuring range of 33 m without intervention on the weights. A wooden beam pinned to the ground and bearing a glued on measuring band with 0.2 cm resolution and $380.0 \pm 0.2$ cm coverage, served as B. The indicator (I) consisted in triangular PMMA plate with a marker line perpendicular to a side, which could be reversibly clamped to $H_2$. Placement of a second indicator at the origin of B, when the first had reached the end, allowed extension of the measuring range.

**Electronics** — The emitter antenna, E, was a naked, streched, center-fed half-wave dipole constructed from two 16.2 cm aluminum tubes of 6 mm diameter. It possessed a standing wave ratio (SWR) of 1.1 at 433.5 MHz. It received its input through a RG58 coaxial cable of length 1100 cm, which close to E formed a current choke balun of 6.5 windings on a 3 cm diameter core made of paper-based laminate. This cable produced a loss of about 3.7 dB. The core of the balun, which enclosed the sliding tube of $P_E$ and the coaxial cable to a length of 14 cm, pointed downward along C. The cable continued to the location of the radio operator near B. A hand-held 5 W, 433.5 MHz ($\lambda$ = 69.16 cm) VHF/UHF transceiver (FT-470, Yaesu Musen Co., Tokyo) fed E. It possessed the two power levels 5.0 W and 0.25 W, i.e. a difference of 13 dB, and it had been provided with a power cable with external switch for connection to a 12 V, 7 A h lead battery. The assembly of R and $P_R$ essentially mirrored that of E and $P_E$, but the aluminum tubes measured 16.7 cm each, which resulted in a SWR of 2.6. A 70 cm long RG58/U coaxial cable connected the R to A along C and through the 5 mm hole in the $P_A$. The electronic measuring components formed a 50-Ω impedance system. The total loss of the system was about 2 dB, of which ca. 0.25 dB originated in the cable, ca. 0.2 dB in the excess SWR,[8] and ca. 1.5 dB in the three connectors.

The ferrite tiles covering PA reduced reflectivity of PA and A for 433.5 MHz radiation at normal incidence by 17 dB.[7] Their efficacy at that frequency is only very weakly dependent on the thickness of the dielectric layer between them and an underlying metal plate, as suggested by data on an almost identical product.[9] The first component of A was a UHF bandpass filter Wisi (Wilhelm Sihn AG, Mägenwil, Switzerland), with the following properties: -3-dB-passband from 425 to 437 MHz, with roll-offs on both sides of 3.4 MHz/decade; insertion loss of 2.225 dB at 432.330 MHz. A very short coaxial cable connected the filter to a DC 500 MHz, 92 dB demodulating logarithmic amplifier (AD8307, Analog Devices Inc., Norwood, MA) with a bandwidth of up to 500 MHz,[10] modified in two ways. The connector to the amplifier received a 50-Ω noninductive resistance and a 5 V supply voltage stabilizer was built into a separate shielded compartment of the amplifier housing.

A 100 m four-strand telephone cable, 0.0883 N/m, hang down from the verticality adjuster of $P_A$ and continued on the ground to a location near B. Through it the modified amplifier, which consumed only 7.5 mA, was powered by a pack of 5 AA batteries that delivered nominally 7.5 V. It further connected the output terminal of the amplifier with a digital multimeter in the hands of the radio operator. Experimenters remained near B during data collection.

**III. THEORY AND DATA ANALYSIS**

The DC output voltage of the amplifier[10], $U_i$, when R is at the distance $z_i$ from E, is linked to the input voltage ratio, $u_i/u_1$, by

$$U_i = U_1 + sL_i = U_1 + 20s\log(u_i/u_1). \tag{1}$$

Therein $s$ = 25 mV/dB, $U_1$ is the output voltage and $u_1$ the sine-amplitude input voltage when R is in the reference position at the distance $z_1$ from E, $u_i$ is the corresponding sine amplitude input voltage, and $L_i$ the input level, when the distance is $z_i$. Only output voltages between 250 mV and 2100 mV can be meaningfully converted into input voltage ratios by Eq. (1). Within the measuring range the output voltage varies with a linearity of ±3 dB with $L_i$. Actually, within the smaller range utilized in the present experiment, the linearity is ±1 dB.

The electric field strength of a half-wave dipole antenna beyond the far field limit diminishes in inverse proportion to the distance. The input voltage sensed by R is proportional to the electric field. This leads to the distance dependence of the input voltage ratio as

$$u_i/u_1 = \exp_{10}[(U_i - U_1) / 500] = (z_1/z_i)^q , \qquad (2)$$

wherein theoretically $q = 1$.

Data were analyzed with the help of the open source statistical software sciDAVis, version created by Qt/QMake, loaded down Apr. 20, 2014. Data fitting involved nonlinear least-square analysis with the help of the Levenberg-Marquardt convergence improvement modification of the Gauss-Newton algorithm, with a tolerance of 0.0001.

## IV. RESULTS AND DISCUSSION

In support of the validity of the new test range, the inverse square law for transmission of 433.5 MHz ($\lambda$ = 69.16 cm) radiation between two aligned naked half-wave dipole antennas was tested. Initially R was placed at $z_1$ = 200 cm from E, i.e. at 1.4 times the $2\lambda$ limit for the far field region of a half-wave dipole antenna. Between $z_i$ = 200 and 1100 cm R was moved stepwise 4 cm at the time, and thereafter 16 cm at the time, up to 3228 cm. The latter distance corresponds to 46 wavelengths. In each position first the instantaneous background multimeter reading, $U_{B,i}$ in mV, was taken. It measured environmental radiation with frequencies centered within the passband of the filter, or centered outside thereof, but tailing into it. Typically about 10 s thereafter the reading with transmission of 433.5 MHz radiation, $U_i$, was recorded. For the interval from 200 cm (1870 mV) to 2604 cm (1281 mV), the transceiver operated at the lower, and from 2604 cm (1571 mV) to 3228 cm (1566 mV) at the higher of its two output power levels. The data were put on the same scale by multiplication of the high power readings with the factor 0.8155 (-12.3 dB), which was determined as the average from paired measurements of $U_i$ with the two power levels in five positions. In several positions measurements were repeated during a backward move of R.

The average reproducibility was 0.3%. Occasional gusts of wind caused R to oscillate visibly. This was accompanied by oscillations in multimeter readings of typically less than 0.3%. Displacement of the telephone cable that connected the amplifier with the multimeter on the ground by several meters did not alter readings.

The distance dependence of the background amplifier output, $U_{B,i}$, showed a constant component of 252 mV inherent to the amplifier, and spikes that in an extreme case attained 641 mV (Fig. 3). The spikes lasted most of the time less than 30 s. In some cases measurements in adjacent positions actually delineated peaks. No relationship between spikes or peaks and local maxima in transmission data, $U_i$, could be detected. The spikes were interpreted as position-dependent alteration in exposure to, and time-of-day-dependent intensity increases in environmental radiation pulses. Two facts argued against a pairing of background and transmission data from the same position of R. These were the short duration of the background spikes relative to the time laps between the two measurements and the lack of knowledge about interference between experimental and environmental radiation.

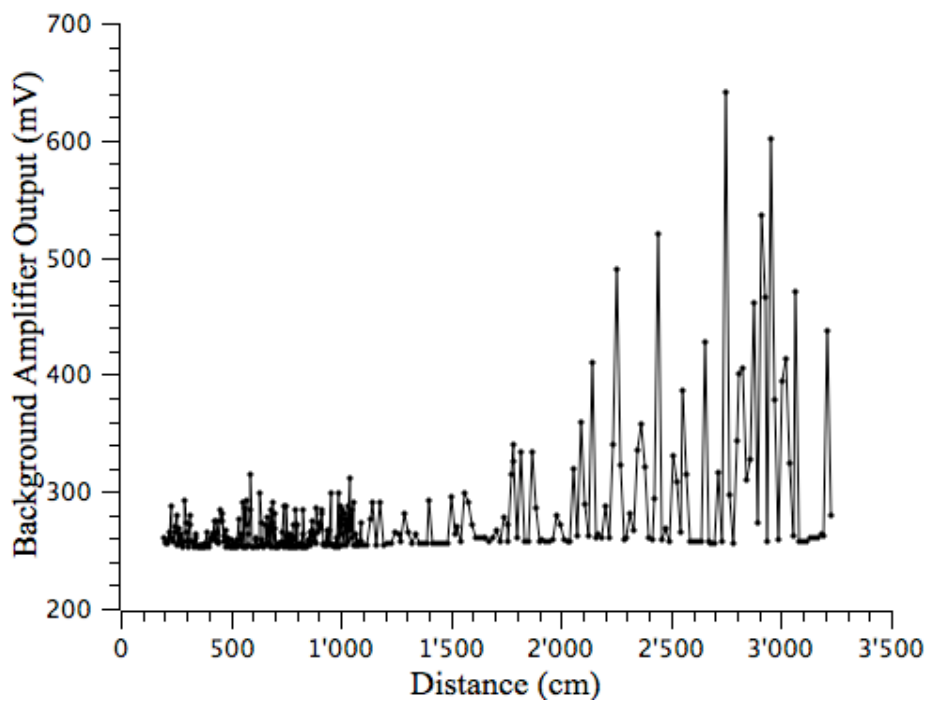

Fig. 3: Background amplifier output, $U_{B,i}$, as a function of the distance between the emitter and receiver antennas, $z_i$.

Correction of transmission data for background was evidently not straightforward. Therefore it was analyzed whether neglect of background could be justified. For this crude analysis the objections against pairing of data was put aside. Figure 4 shows the dependence on $z_i$ of the relative background input, $u_{B,i,rel}$, i.e. the % of background input voltage, $u_{B,i}$, relative to the transmission input voltage, $u_i$, according to the equation

$$u_{B,i,rel} = 100(u_{B,i}/u_i) = 100\exp_{10}[(U_{B,i} - U_i) / 500] , \qquad (3)$$

Up to $z_i = 1756$ cm the $u_{B,I,rel}$ maximally reached 0.6%, and in most of the positions it remained actually much below. These values were considered negligible. At larger distances $u_{B,i,rel}$ increased, in the worst case to 2.5%.

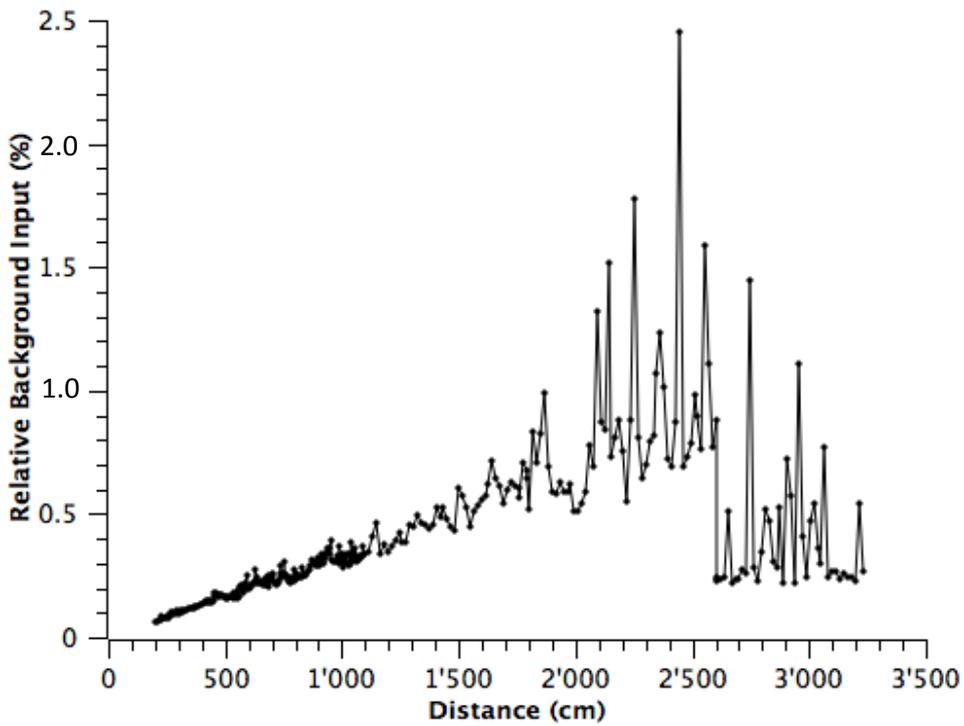

Fig. 4: Relative background input, $u_{B,i,rel}$, as a function of the distance between the emitter and receiver antennas, $z_i$. At 2604 cm the power of the transceiver was switched from 0.25 to 5 W.

A glance at $U_i$ plotted against $\log z_i$ revealed the straight line expected from Eq. (2) and a slope consistent with $q = 1$. Of note were frequent data from adjacent positions deviating in the same direction from the line, immediate evidence of non-randomness of deviations. Note that recognition of the non-randomness was only made possible by the elevated spatial resolution of the measurements. Anticipating the final analysis, deviations of up to 27% may also be seen in Fig. 5. Their magnitude excluded measurement error and background radiation as their origin. Experimental radiation had to be blamed. The deviation's stationary character indicated standing waves.

Sets of deviations were computed using for the description of the general course of the distance dependence Eq. (2) and assuming various plausible parameter values. Fourier analyses were performed on the entire sets of deviations and on subsets thereof. The spectra always showed numerous maxima. In no case did a Fourier spectroscopic wave number stand out. In particular, none related to the experimental wavelength or its half accounted ever for more than 0.9% of the sum of all Fourier amplitudes. This excluded an equipment internal origin of the variations.

It could be calculated that ground reflection on the present test range allowed a constructive interference signal only from radiation with a small backward component. At best a single broad signal maximum at around $z_i = 932$ cm, corresponding to a phase difference of $2\lambda$, could be expected. No sign of such a maximum was evident. Thus ground reflection was too weak to explain position and sharpness of any of the deviation peaks. The variations became maximal roughly in the middle of the test stretch. Proximity to the tower did apparently not play a dominant role in their magnitude. Since no specific sources of variations within the equipment could be identified, it had to be concluded that they originated in complex interferences between the direct experimental radiation beam and reflections thereof from a multitude of objects in the environment, such as trees and components of the viewing tower.

Testing the applicability of the inverse square law involved the evaluation of the value of $q$ in Eq. (2), which theoretically is 1. Because of the manner $z_i$ was read off B, measurement errors in consecutive data could not accumulate and became actually compensated. For analytical purposes this allowed to consider the error in $z_i$ negligible. The measured electronic signal thus was the observable with the dominant error. Regression analysis requires a mathematical model function from whose predictions the data deviate minimally according to a chosen criterion and the deviations reflect random error. The mentioned non-randomness of the deviations called for a model function beyond the inverse square law, but mathematical modeling of the situation in its full complexity was unrealistic. The multitude of interference sources let one expect that summation of constructive and destructive interferences resulted in positive and negative deviations with very similar probabilities and amplitudes. In this respect such deviations resemble measurement error, being distinct mainly through the non-randomness of distribution of signs. These properties suggested that a classical regression analysis would produce a fit to the law with a balanced distribution of the deviations, but without statistical rigor. The extent to which under such circumstances the error estimates on parameters maintain their usual meaning is questionable. Since the coefficient of determination, $R^2$, is independent of the fitting method and convergence criterion, and the inverse square law described the general course of the data well, it served here as the decisive optimization parameter.

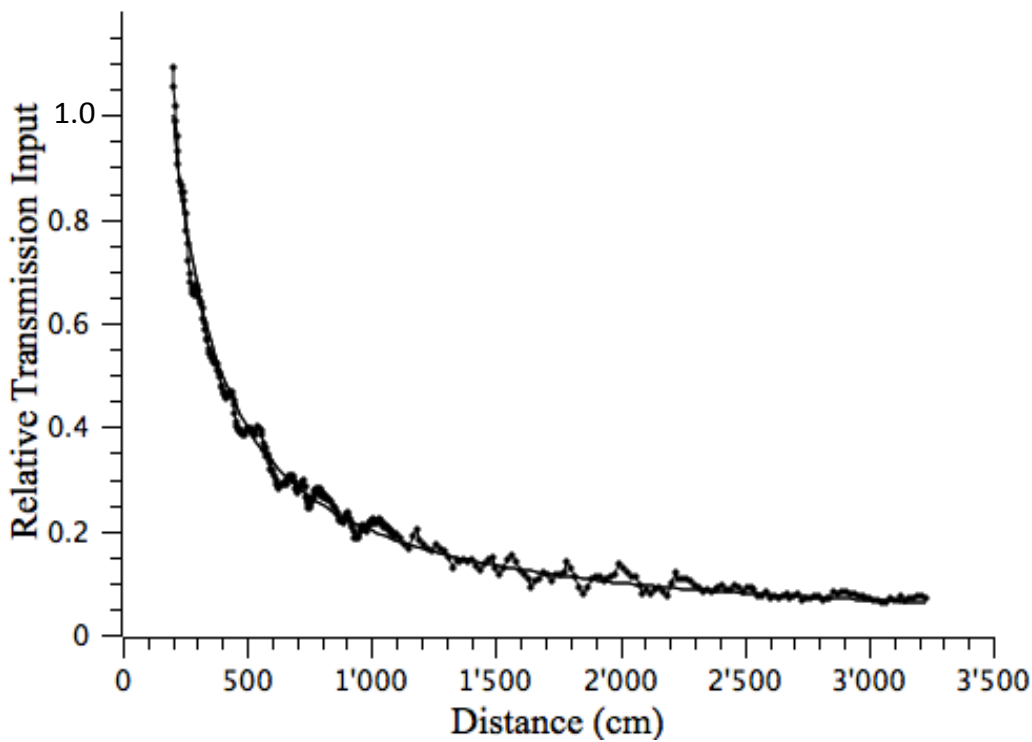

Fig. 5: Relative transmission input (= input voltage ratio) at the receiver antenna as a function of the distance between the emitter and receiver antennas, according to Eq. (2). Data are individual measurements, normalized to the lower power level of the transceiver, as described in the text. The solid line is the least square fit given by the equation for the input voltage ratio $u_i/u_1 = \exp_{10}[(U_i - 1851)/500] = 200/z_i$.

Among various approaches to fitting Eq. (2), the following one yielded the fit with the highest $R^2$ value. Various values for $U_1$ in the vicinity of a visual estimate were used to compute through Eq. (2) sets of values of $u_i/u_1$. Each data set was then fit by Eq. (2), with $z_1$ and $q$ as adjustable parameters. The best combination of parameter values turned out to be: $U_1 = 1851$ mV, $z_1 = 199.59 \pm 0.90$ cm, $q = 0.9970 \pm 0.0051$, with $R^2 = 0.992$ (Fig. 5). Elimination of the data with $z_i$ larger than 1756 cm, among which some particularly large background spikes existed, did not improve the fit. This supported neglect of the background. In order to exclude rigorously any residual near field effects of the emitter antenna on the distance dependence, the data subset from distances beyond 10 wavelengths, i.e. > 7 m, were similarly analyzed. The result confirmed what was found with the full data set. Thus in the present endeavor near field effects played no role.

In summary, the best-fit $z_1$ matched well the measured closest inter-antenna distance of 200 cm, helping to validate the analysis. The exponent $q$ turned out indistinguishable from the theoretical value of 1. The applicability of the inverse square law vouched for the suitability of the test range design for distance studies. Post hoc the results supported in the particular case the treatment of non-random deviations as surrogate random errors. Given the origin of the deviations, the procedure should remain valid also for other experiments performed on a test range with similar limitations.

The analysis of deviations suggests that the use of a strictly metal-free tower in a completely flat and tree-free environment would reduce or even completely eliminate the undesirable variations. But even under present conditions the antenna test range should allow extension of the studies to other questions, e.g. the simultaneous variation of distance and orientation of various kinds of receiver antennas. The novel test range allows repeated and highly reproducible movement of R between any chosen measurement position and the bottom one. In the latter position R on an appropriate P can easily be given a particular orientation. The transmission data of interest are obtainable through regression analysis of the distance dependence around the distance of choice.

## I. CONCLUSION

In conclusion, a novel, open-air, variable-distance antenna test range has been built, which offered elevated spatial resolution and satisfied the requirements for studies of transmission of 433.5 MHz radiation under quasi free-space conditions. Except for the tower, the equipment is easily affordable for most undergraduate physics departments. Equipment-internal interferences could be completely avoided and means for dealing with residual interferences with the environment were found. The performance of the range was evaluated through examination of the inverse square law in the case of a pair of parallel naked half-wave dipole antennas. Beyond a distance 1.4 times the $2\lambda$ limit for the far field region, the ratio of sine amplitude input voltages to the receiver antenna at two different distances between emitter and receiver antennas diminished in proportion to the corresponding inverse distance ratio to the power $0.9970 \pm 0.0051$ ($R^2 = 0.992$), reflecting the behavior of the ratio of electric field strengths. The value of the exponent is indistinguishable from the theoretical value of 1. Given the known proportionality of irradiance to the square of the electric field strength, the result corroborates the inverse square law for irradiance. The corroboration regards the lowest frequency for which such studies have been published.

ACKNOWLEDGMENTS

The authors thank Mr. Andres Ammann, and Verein Lysser Aussichtsturm Personalwaldkorporation Lyss for the permission to perform experiments at their viewing tower.

___________________________________________